# Stabilization of vector solitons in optical lattices

Yaroslav V. Kartashov,[1,2] Anna S. Zelenina,[2] Victor A. Vysloukh,[3] Lluis Torner[1]

*[1]ICFO-Institut de Ciencies Fotoniques, and Department of Signal Theory and Communications, Universitat Politecnica de Catalunya, 08034 Barcelona, Spain*

*[2]Physics Department, M. V. Lomonosov Moscow State University, 119899, Vorobiovy Gory, Moscow, Russia*

*[3]Departamento de Fisica y Matematicas, Universidad de las Americas - Puebla, Santa Catarina Martir, CP 72820, Puebla, Cholula, Mexico*

We address the properties and dynamical stability of one-dimensional vector lattice solitons in Kerr-type cubic medium with harmonic transverse modulation of refractive index. We discovered that unstable families of scalar lattice solitons can be stabilized via the cross-phase modulation (XPM) in the vector case. It was found that multi-humped vector solitons that are unstable in uniform media where XPM strength is higher than that of self-phase modulation, can also be stabilized by the lattice.

**1. Introduction**. Light propagation in media whose properties vary periodically in transverse direction exhibits a wealth of practically interesting phenomena including the formation of stable localized structures, which find applications in many branches of modern physics, including waves in molecular chains [1], trapped Bose-Einstein condensates [2], or solids [3]. In nonlinear optics discrete solitons were extensively studied and observed in periodic arrays of weakly coupled waveguides [4]. Such strongly localized modes might be used to test all-optical switching and routing concepts. Recently it was shown that lattices constituted by continuous nonlinear media with an imprinted harmonic modulation of the refractive index offer a number of additional opportunities for the manipulation of light signals [5]. Scalar solitons in the harmonic lattices were observed in photorefractive crystals [6] and analyzed in Refs [7-8].

However, the interaction between several light waves can considerably enrich the dynamics of their propagation and open new perspectives for cross-stabilization and all-optical soliton phenomena. In uniform media, two-component bright vector solitons were studied for coherent [9] and incoherent [10] interactions. Strongly localized vectorial *discrete modes* in arrays of evanescently coupled waveguides and their stability were reported in Ref. [11]. Very recently coupling between mutually incoherent solitons belonging to the different bands of transmission spectrum of periodic lattices was discussed [12], while vector solitons were observed in AlGaAs nonlinear waveguide arrays [13]. However, the investigation of the properties and stability of *vector solitons in optical lattices*, which can be qualitatively and quantitatively altered by variation of the properties of the lattice, is an open problem.

In this work we perform a detailed analysis of the properties and dynamical stability of one-dimensional vector lattice solitons in both focusing and defocusing cubic media. We reveal that cross-phase-modulation (XPM) results in stabilization of even (in focusing medium) and twisted (in defocusing medium) soliton components that are known to be highly unstable in the scalar case. We show that, in contrast to solitons in uniform media, vector lattice solitons can be made stable if the XPM strength exceeds that of self-phase modulation. We also reveal the existence of stable multi-humped vector complexes in which one component is stabilized by the lattice, while the other one is stabilized by XPM.

**2. Model.** We address the propagation of coupled laser beams along the $\xi$-axis in media with a periodic modulation of the linear refractive index in the $\eta$-direction, described by the system of coupled nonlinear Schrödinger equations:

$$\begin{aligned} i\frac{\partial q_1}{\partial \xi} &= -\frac{1}{2}\frac{\partial^2 q_1}{\partial \eta^2} + \sigma q_1(|q_1|^2 + C|q_2|^2) - pR(\eta)q_1, \\ i\frac{\partial q_2}{\partial \xi} &= -\frac{1}{2}\frac{\partial^2 q_2}{\partial \eta^2} + \sigma q_2(C|q_1|^2 + |q_2|^2) - pR(\eta)q_2, \end{aligned} \tag{1}$$

where $\eta$ and $\xi$ are scaled to the beam width and diffraction length, respectively; $C$ is the XPM parameter; $\sigma = \mp 1$ for the focusing/defocusing medium; $p$ is the guiding parameter; function $R(\eta) = \cos(2\pi\eta / T)$ describes refractive index profile, and $T$ is the lattice period. The XPM coefficient in Eqs. (1) depends on the particular settings and materials involved. Thus for mutually incoherent light beams $C=1$ [10,14], while for coherent orthogonally polarized beams interacting in the highly birefringent media $C=2$ [9,15]. The parameter $C$ can acquire quite high values in organic materials [9]. Notice also that photorefractive crystals offer new possibilities for manipulation of the XPM coupling, by varying the polarization of the light beams or the elements of electro-optic tensor involved [6]. Eqs (1) admit several conserved quantities including the total $U$ and partial $U_{1,2}$ energy flows:

$$U = U_1 + U_2 = \int_{-\infty}^{\infty} (|q_1|^2 + |q_2|^2) d\eta. \quad (2)$$

**3. Vector soliton families**. Stationary solutions of Eqs (1) have the form $q_{1,2}(\xi,\eta) = w_{1,2}(\eta)\exp(ib_{1,2}\xi)$, where $w_{1,2}(\eta)$ are real functions and $b_{1,2}$ are real propagation constants. Lattice soliton families are defined by $b_{1,2}$, the lattice period $T$, and parameters $p$ and $C$. Since one can use scaling transformation $q_{1,2}(\eta,\xi,p,C) \to \chi q_{1,2}(\chi\eta,\chi^2\xi,\chi^2 p,C)$ to obtain various families of solutions from a given one, the transverse scale was selected in such way that modulation period $T=\pi/2$ is a constant, and $b_{1,2}$, $p$, $C$ are variable parameters. Upon linear stability analysis we searched for the perturbed solutions of Eqs (1) in the form $q_{1,2}(\eta,\xi) = [w_{1,2}(\eta) + u_{1,2}(\eta,\xi) + iv_{1,2}(\eta,\xi)]\exp(ib_{1,2}\xi)$, where real $u_{1,2}$ and imaginary $v_{1,2}$ parts of small perturbation can grow with complex growth rate $\delta$. The standard linearization procedure around stationary solution $w_{1,2}$ for Eqs (1) yields the linear eigenvalue problem

$$\begin{aligned}
\delta u_1 &= -\frac{1}{2}\frac{d^2 v_1}{d\eta^2} + [\sigma(w_1^2 + Cw_2^2) + b_1]v_1 - pRv_1, \\
\delta v_1 &= \frac{1}{2}\frac{d^2 u_1}{d\eta^2} - [\sigma(3w_1^2 + Cw_2^2) + b_1]u_1 - 2\sigma C w_1 w_2 u_2 + pRu_1, \\
\delta u_2 &= -\frac{1}{2}\frac{d^2 v_2}{d\eta^2} + [\sigma(w_2^2 + Cw_1^2) + b_2]v_2 - pRv_2, \\
\delta v_2 &= \frac{1}{2}\frac{d^2 u_2}{d\eta^2} - [\sigma(3w_2^2 + Cw_1^2) + b_2]u_2 - 2\sigma C w_1 w_2 u_1 + pRu_2,
\end{aligned} \quad (3)$$

for the perturbation components $u_{1,2}, v_{1,2}$ that was solved numerically. Scaling transformation mentioned above predicts changes of the growth rate for unstable solitons with identical functional profiles supported by lattices with different periods. For example, if lattice period becomes $\chi$ times smaller then respective growth rate increases $\chi^2$ times.

We start our analysis by recalling the properties of *scalar* lattice solitons. There exist odd, even, and twisted scalar lattice solitons. In focusing media odd solitons are stable, even ones are unstable, and twisted ones are stable above certain energy threshold. Defocusing media supports stable odd and even solitons, but twisted solitons are unstable. The simplest vector soliton solutions are formed at $b_1 = b_2$, when $w_1(\eta) = w(\eta)\cos\phi$, $w_2(\eta) = w(\eta)\sin\phi$, where $w(\eta)$ is the scalar soliton profile, and $\phi$ is a phase. The most interesting situation occurs at $b_1 \le b_2$, when the first and the second component have different types of symmetry. Below we focus on the simplest self-sustained structures, having twisted first component $w_1$. Such vector solitons can be classified according to field distribution in the second component.

**A. Odd solitons in focusing media**. The properties of *odd* vector soliton in focusing media are summarized in Fig. 1 at $C = 1$. At low energy flows the second component of odd soliton has a single well-defined maximum coinciding with the local maximum of $R(\eta)$, so that the field distribution in both second and first components is asymmetric (Fig. 1(b)). There exist lower and upper cut-offs on $b_1$ at fixed $b_2$ and $p$ (Fig. 1(a)). As $b_1$ approaches upper cut-off, the second component develops two equal humps located on neighboring lattice sites, it means that the odd vector soliton transforms into the even one. At lower cut-off odd soliton ceases to exist. Energy flow versus $b_1$ is shown in Fig. 1(a) at fixed $b_2$ and $p$. Energy flow drops off with growth of guiding parameter at fixed $b_1$ and $b_2$. At $U, U_{1,2} \to \infty$ odd vector soliton transforms into weakly coupled scalar and vector solitons located at neighboring sites. The area of existence of odd soliton first expands and then shrinks with growth of guiding parameter $p$ (Fig. 1(c)) for fixed $b_2$, so odd solitons cease to exist, when the guiding parameter $p$ exceeds the critical value. The width of existence area increases with growth of $b_2$. Notice that there are lower threshold on $b_2$ for existence of odd solitons.

**B. Even solitons in focusing media**. The properties of *even* vector solitons (composed from twisted first and even second components) are summarized in Fig. 2.

Second component has two equal intensity maxima located on neighboring sites (Fig. 2(b)). Total energy flow decreases monotonically with growth of $b_1$ at fixed $b_2$, $p$, and drops off with increase of $p$ at fixed $b_1$, $b_2$ (Fig. 2(a)). There are lower and upper cut-offs on $b_1$. At lower cut-off $w_1 \to 0$, and vector soliton transforms into even scalar soliton, while at upper cut-off $w_2 \to 0$, and it converts into twisted scalar soliton. The existence area of even soliton expands with decrease of guiding parameter $p$ and slightly changes with growth of $b_2$ (Fig. 2(c)).

**C. Soliton stability in focusing media**. Results of stability analysis are summarized in Figs. 1, 2. We have found that *odd* vector solitons are stable almost in the whole domain of their existence (Fig.1 (c)), a result confirmed by numerical integration of Eqs (1) in the presence of input noise (Fig. 1(d)). The linear stability analysis also revealed existence of stability bands for *even* vector solitons. They turn to be stable when the amplitude of first twisted component becomes large enough. Therefore XPM can stabilize the otherwise unstable soliton component. This is one of the most important results of this work.

Soliton stabilization occurs because of local increase of refractive index in neighboring lattice sites created by stable twisted component via XPM. This local increase prevents even component from decay into odd one under action of perturbations. The onset of stability is dictated by peak amplitude or energy $U_1$ of twisted component, ratio $U_1/U_2$, and depth of optical lattice that can be flexibly controlled in distinction from discrete systems. Since lower amplitudes are necessary to support soliton-like propagation in deeper lattices, stabilizing action of twisted component via XPM and width of stability domain decreases with growth of $p$. Notice that upper boundary of instability domain for even vector soliton coincides with upper cut-off for odd one, i.e. latter transform into stable even soliton at upper cut-off. Fig. 2(d) illustrates stable propagation of even lattice soliton perturbed by noise.

**D. Odd and even solitons in defocusing media**. Optical lattices in defocusing media also support vector solitons, but those are typically wider than their counterparts in focusing media. The energy flow of *even* soliton versus $b_1$ is depicted in Fig. 3(a), while Fig. 3(b) shows profile of such soliton. The energy flow increases with growth of $p$ at fixed $b_1, b_2$. Notice that in defocusing media at upper cut-off on $b_1$, the first component vanishes and vector soliton transforms into even scalar one, while at lower cut-off on $b_1$ second component vanishes and one gets twisted scalar soliton. The area of existence for even solitons at $(b_1, p)$ plane broadens with decrease of guiding parameter (Fig. 3(c)) at fixed $b_2$. The width of existence area on $b_1$ decreases linearly with growth of propagation constant $b_2$ at fixed $p$, so that above certain threshold on $b_2$ even solitons cease to exist. We have also found *odd* vector solitons in defocusing medium (Fig. 3(e)). Its first component transforms into linear Bloch wave at the lower cut-off on $b_1$, while the second one remains localized. Odd soliton converts into the stable even soliton at upper cut-off on $b_1$.

**E. Soliton stability in defocusing media**. Results of stability analysis are summarized in Fig. 3(c). We discovered existence of stability band for even vector solitons. Thus in defocusing optical media twisted first component (that is unstable alone) can be stabilized through XPM interaction with the stable even second component. The cross-stabilization takes place if the amplitude of second even component is large enough, actually near the upper cut-off for existence. Stability area for even vector solitons shrinks at low $p$. Fig. 3(d) shows stable propagation of even soliton in defocusing media in the presence of white input noise. Stability analysis for odd solitons becomes complicated near lower cut-off (area of weak localization) but we have found that they are stable near upper edge of existence domain on $b_1$ (Fig. 3(f)). The important result is that combined action of the lattice and XPM enables to stabilize vector solitons of *high order,* to be perhaps referred to as *vector soliton trains,* with complex multi-humped intensity profile, when one of component is a scalar soliton train.

**F. Impact of XPM strength on soliton stability**. We also analyzed the impact of XPM strength on stability of vector lattice solitons. The most important result is that stability window for bright even lattice solitons exists at $C \neq 1$ (Fig. 4(a)) in contrast to the case of uniform media, where multi-humped vector solitons are unstable at $C > 1$ in the entire domain of their existence, and the width of stability window can be increased with increase of lattice depth. Notice that even soliton component vanishes at lower cut-off on $b_1$ at $C \gtrsim 1.03$ thus resulting in stabilization of vector soliton, while at $C < 1.03$ even component vanishes at upper cut-off (Fig. 4(a)).

Stabilization via XPM is also possible when first soliton component is subject to influence of lattice ($p \neq 0$) while for the second component the medium is uniform ($p = 0$). It was revealed that XPM may results in completely stable twisted-twisted soliton combination (Fig. 4(b)) that does not exist in uniform medium. In this case even weak first “lattice” component can capture and stabilize strong second “uniform” component. Existence and stability domains broaden with growth of lattice depth in first component (Fig. 4(c)). Because of alternating sign of “lattice” component in neighboring sites, the “uniform” component also acquires multi-humped structure (Fig. 4(e)), but still shows stable propagation (Fig. 4(f)).

**4. Conclusions.** In conclusion, we analyzed the properties of vector lattice solitons in cubic nonlinear media with harmonic transverse modulation of linear refractive index, and discovered that stable propagation sustained by XPM if possible even if one of the soliton components is otherwise unstable. Here we reported only the simplest examples of vector lattice solitons but results are expected to hold for more general settings and for richer field distributions, including the formation of stable soliton trains. We stress that the combined action of the lattice with tunable strength and XPM offers opportunities not only to alter the *quantitative* characteristics of solitons, but also to control their *qualitative* features, including their topological structure and stability.

This work has been partially supported by the Generalitat de Catalunya, and by the Spanish Government through grant BFM2002-2861.

## References

1. A. S. Davydov and N. I. Kislukha, Phys Status Solidi B **59**, 465 (1973).
2. A. Trombettoni and A. Smerzi, Phys. Rev. Lett. **86**, 2353 (2001).
3. W. P. Su et al., Phys. Rev. Lett. 42, 1698 (1979).
4. D. N. Christodoulides and R. I. Joseph, Opt. Lett. **13**, 794 (1988); D. N. Christodoulides et al., Nature **424**, 817 (2003); H. S. Eisenberg et al., Phys. Rev. Lett. **81**, 3383 (1998); R. Morandotti et al., Phys. Rev. Lett. **83**, 2726 (1999).
5. Y. V. Kartashov et al., Opt. Lett. **29**, 766 (2004); **29**, 1102 (2004); Opt. Express **12**, 2831 (2004).
6. N. K. Efremidis et al., Phys. Rev. E **66**, 046602 (2002); J. W. Fleischer et al., Phys. Rev. Lett. **90**, 023902 (2003); Nature **422**, 147 (2003); D. Neshev et al., Opt. Lett. **28**, 710 (2003).
7. N. K. Efremidis et al., Phys. Rev. Lett. **91**, 213906 (2003).
8. P. J. Y. Louis et al., Phys. Rev. A **67**, 013602 (2003); N. K. Efremidis and D. N. Christodoulides, Phys. Rev. A **67**, 063608 (2003).
9. V. Manakov, Sov. Phys. JETP **38**, 248 (1974); D. N. Christodoulides and R. I. Joseph, Opt. Lett. **13**, 53 (1988); J. U. Kang et al., Phys. Rev. Lett. **76**, 3699 (1996); N. Akhmediev et al., Phys. Rev. Lett. **81**, 4632 (1998).
10. M. Mitchell et al., Phys. Rev. Lett. **80**, 4657 (1998); G. I. Stegeman and M. Segev, Science **286**, 1518 (1999); E. A. Ostrovskaya et al., Phys. Rev. Lett. **83**, 296 (1999); C. Cambournac et al., Phys. Rev. Lett. **89**, 083801 (2002).
11. S. Darmanyan et al., Phys. Rev. E **57**, 3520 (1998).
12. O. Cohen et al., Phys. Rev. Lett. **91**, 113901 (2003); A. A. Sukhorukov and Y. S. Kivshar, Phys. Rev. Lett. **91**, 113902 (2003).
13. D. Mandelik et al., Phys. Rev. Lett. **90**, 253902 (2003); J. Meier et al., Phys. Rev. Lett. **91**, 143907 (2003).
14. M. Mitchell et al., Phys. Rev. Lett. **79**, 4990 (1997).
15. C. R. Menyuk, IEEE Journal of Quantum Electronics **23**, 174 (1987); **25**, 2674 (1989).

## Figure captions

Figure 1. (a) Energy flow of odd soliton versus propagation constant $b_1$ at $b_2 = 3$ and various guiding parameters. (b) Profile of odd soliton at $b_1 = 0.7$, $b_2 = 1.5$, $p = 2$. (c) Areas of stability and instability (shaded) for odd solitons on $(b_1, p)$ plane at $b_2 = 3$. (d) Stable propagation of odd soliton depicted in (b) in the presence of white noise with variance $\sigma_{1,2}^2 = 0.01$. In (d) only second component is shown. Focusing medium $\sigma = -1$, $C = 1$. All quantities are plotted in arbitrary dimensionless units.

Figure 2. (a) Energy flow of even soliton versus propagation constant $b_1$ at $b_2 = 3$ and various guiding parameters. (b) Profile of even soliton at $b_1 = 1.65$, $b_2 = 2$, $p = 2$. (c) Areas of stability and instability (shaded) for even solitons on $(b_1, p)$ plane at $b_2 = 3$. (d) Stable propagation of even soliton depicted in (b) in the presence of white noise with variance $\sigma_{1,2}^2 = 0.01$. In (d) only second component is shown. Focusing medium $\sigma = -1$, $C = 1$. All quantities are plotted in arbitrary dimensionless units.

Figure 3. (a) Energy flow of even soliton versus propagation constant $b_1$ at $b_2 = -1$ and various guiding parameters. (b) Profile of even soliton at $b_1 = -1.36$, $b_2 = -1$, $p = 5$. (c) Areas of stability and instability (shaded) for even solitons on $(b_1, p)$ plane at $b_2 = -1$. (d) Stable propagation of even soliton depicted in (b) in the presence of white noise with variance $\sigma_{1,2}^2 = 0.01$. (e) Profile of odd soliton at $b_1 = -1.6$, $b_2 = -1$, $p = 5$. (f) Stable propagation of odd soliton depicted in (e) in the presence of white noise with variance $\sigma_{1,2}^2 = 0.01$. In (d) and (f) only first component is shown. Defocusing medium $\sigma = 1$, $C = 1$. All quantities are plotted in arbitrary dimensionless units.

Figure 4. (a) Areas of stability and instability (shaded) for even solitons on $(C, b_1)$ plane at $b_2 = 3$, $p = 2$. (b) Profile of twisted soliton in the uniform medium supported by twisted lattice soliton at $b_1 = 3.8$, $b_2 = 2$, $p = 4$,

and (c) areas of stability and instability (shaded) for such solitons on the $(b_1, p)$ plane at $b_2 = 2$. (d) Stable propagation of soliton depicted in (b) in the presence of white noise. (e) Profile of three-humped soliton in the uniform medium supported by twisted lattice soliton at $b_1 = 3.57$, $b_2 = 2$, $p = 4$, and (f) its stable propagation in the presence of white noise. In (d) and (f) only second component is shown. Focusing medium $\sigma = -1$. All quantities are plotted in arbitrary dimensionless units.

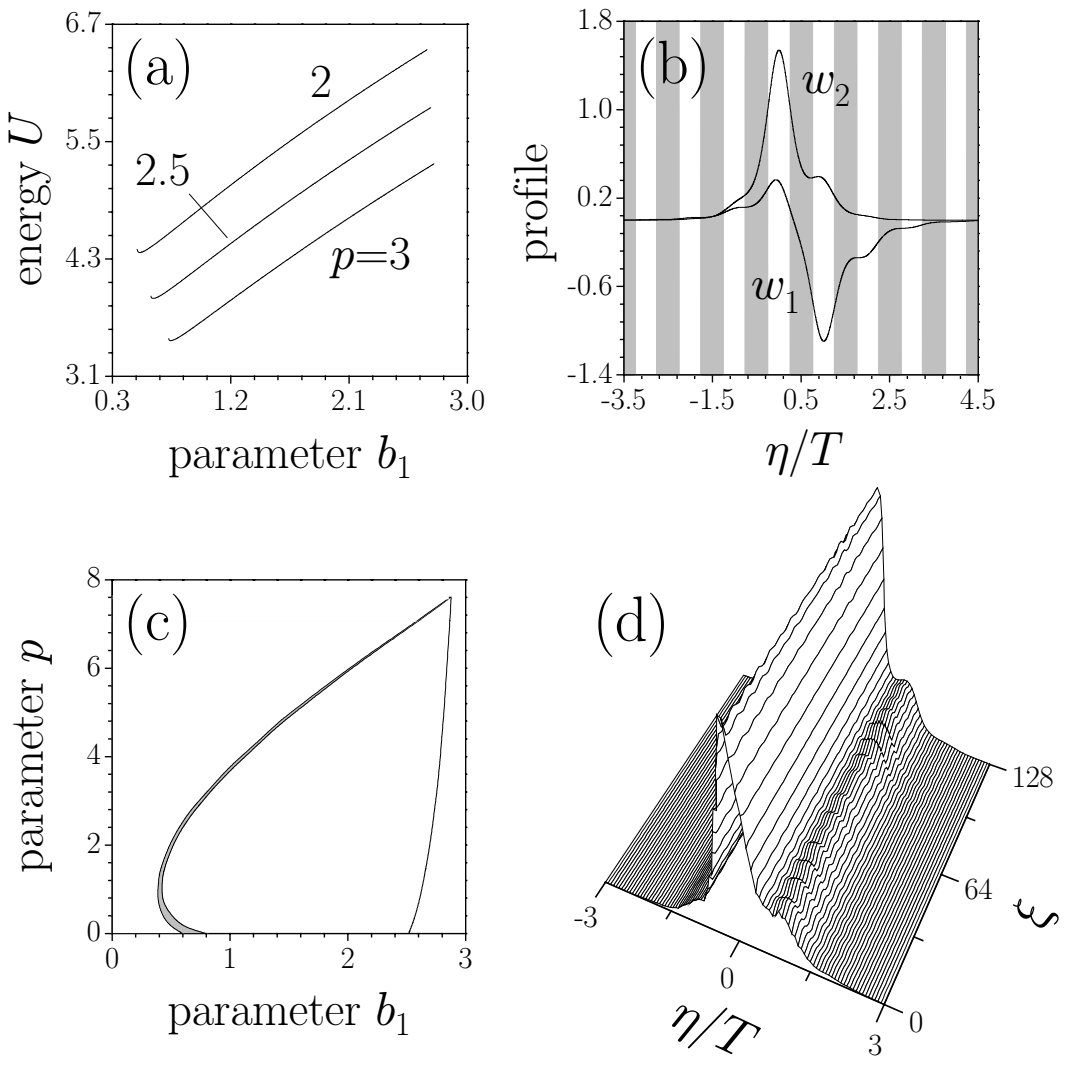

Figure 1. (a) Energy flow of odd soliton versus propagation constant $b_1$ at $b_2 = 3$ and various guiding parameters. (b) Profile of odd soliton at $b_1 = 0.7$, $b_2 = 1.5$, $p = 2$. (c) Areas of stability and instability (shaded) for odd solitons on $(b_1, p)$ plane at $b_2 = 3$. (d) Stable propagation of odd soliton depicted in (b) in the presence of white noise with variance $\sigma_{1,2}^2 = 0.01$. In (d) only second component is shown. Focusing medium $\sigma = -1$, $C = 1$. All quantities are plotted in arbitrary dimensionless units.

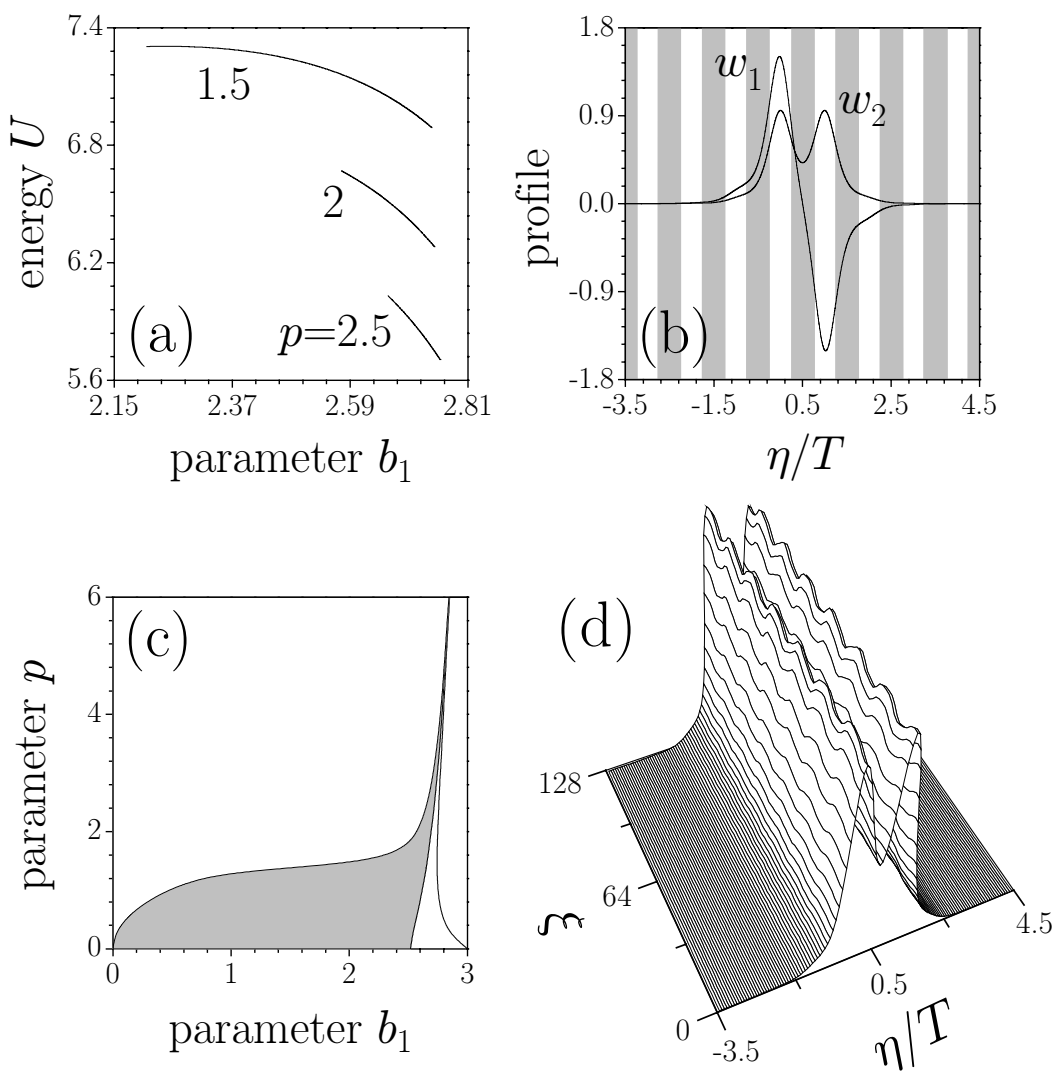

Figure 2. (a) Energy flow of even soliton versus propagation constant $b_1$ at $b_2 = 3$ and various guiding parameters. (b) Profile of even soliton at $b_1 = 1.65$, $b_2 = 2$, $p = 2$. (c) Areas of stability and instability (shaded) for even solitons on $(b_1, p)$ plane at $b_2 = 3$. (d) Stable propagation of even soliton depicted in (b) in the presence of white noise with variance $\sigma_{1,2}^2 = 0.01$. In (d) only second component is shown. Focusing medium $\sigma = -1$, $C = 1$. All quantities are plotted in arbitrary dimensionless units.

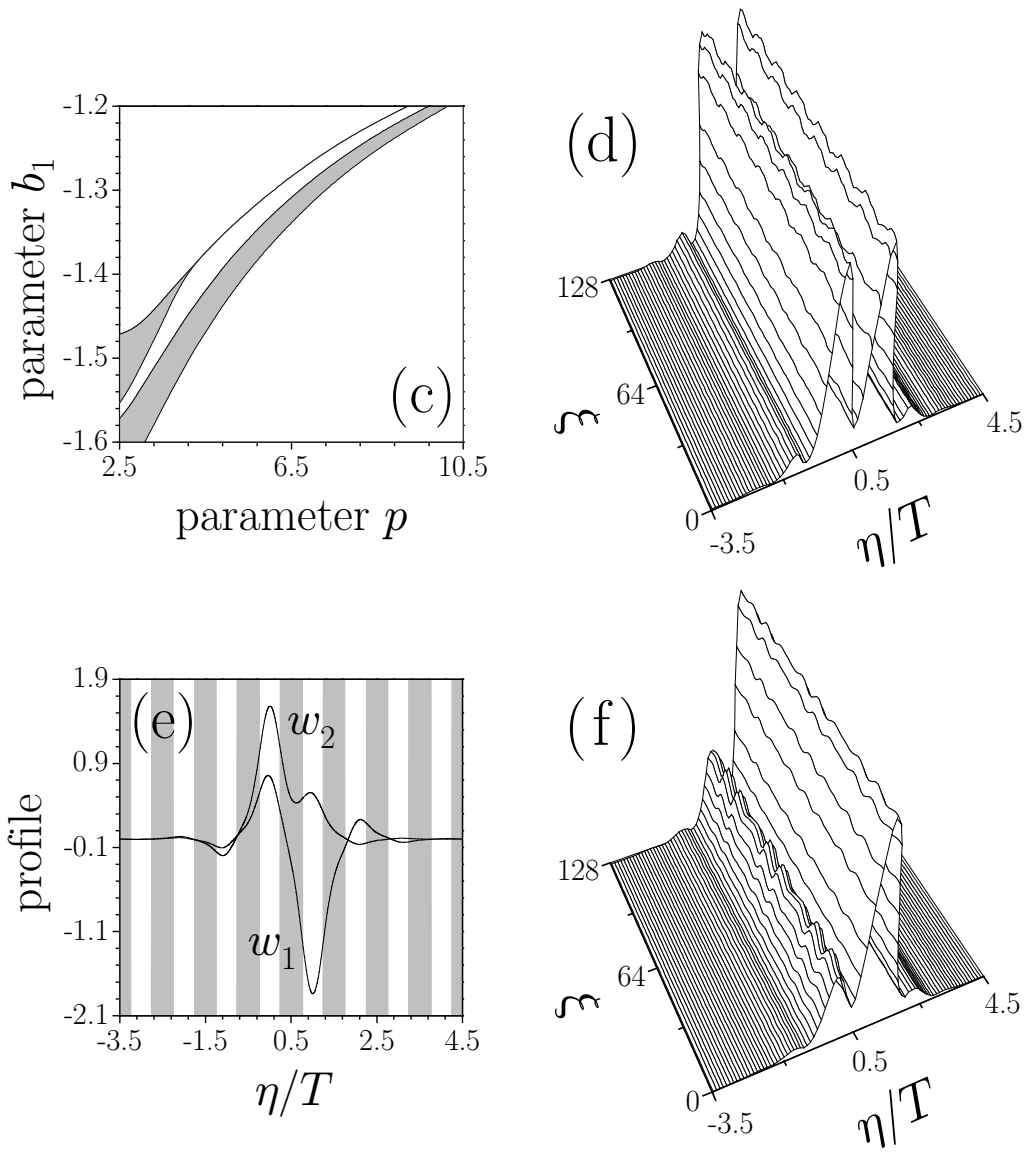

Figure 3. (a) Energy flow of even soliton versus propagation constant $b_1$ at $b_2 = -1$ and various guiding parameters. (b) Profile of even soliton at $b_1 = -1.36$, $b_2 = -1$, $p = 5$. (c) Areas of stability and instability (shaded) for even solitons on $(b_1, p)$ plane at $b_2 = -1$. (d) Stable propagation of even soliton depicted in (b) in the presence of white noise with variance $\sigma_{1,2}^2 = 0.01$. (e) Profile of odd soliton at $b_1 = -1.6$, $b_2 = -1$, $p = 5$. (f) Stable propagation of odd soliton depicted in (e) in the presence of white noise with variance $\sigma_{1,2}^2 = 0.01$. In (d) and (f) only first component is shown. Defocusing medium $\sigma = 1$, $C = 1$. All quantities are plotted in arbitrary dimensionless units.

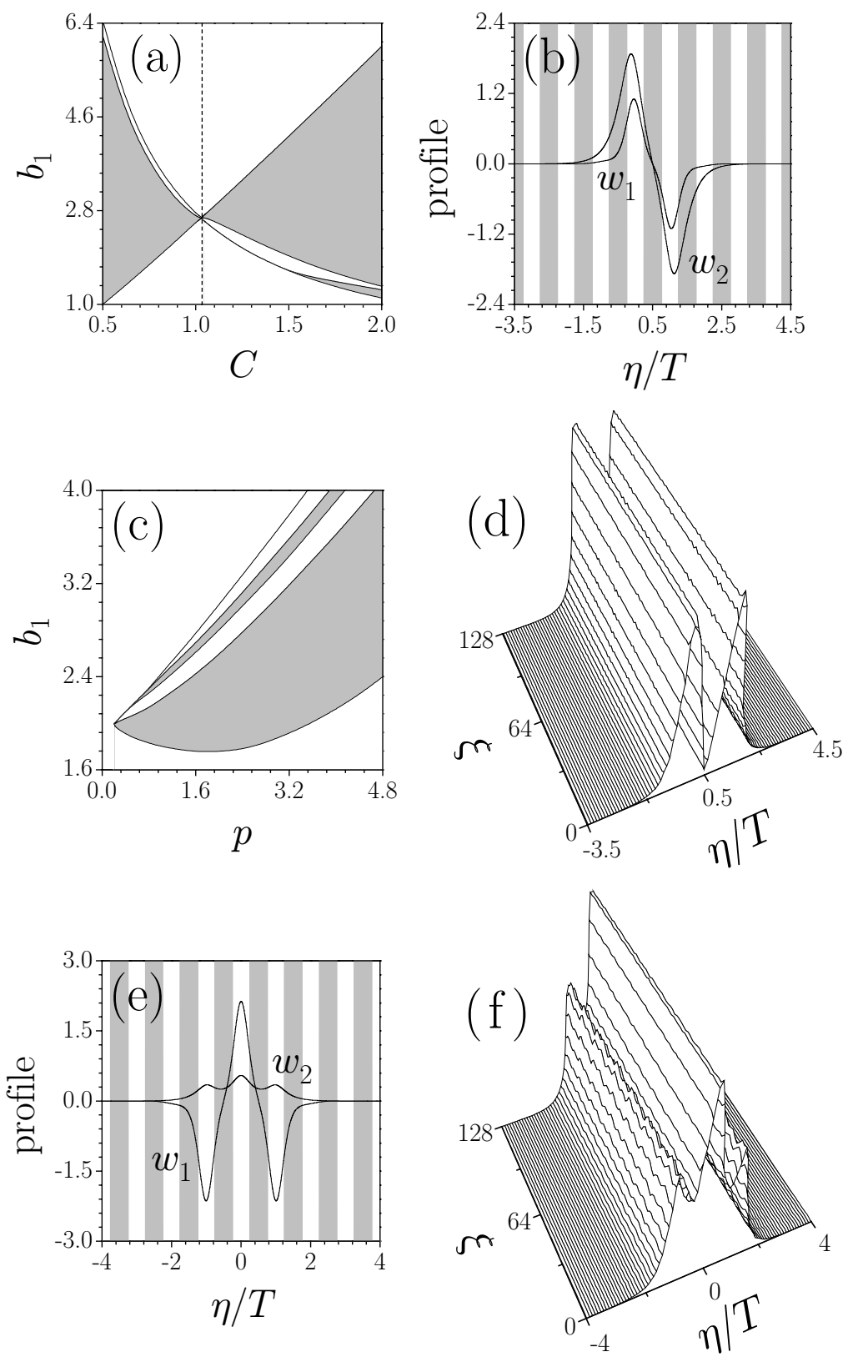

Figure 4. (a) Areas of stability and instability (shaded) for even solitons on $(C, b_1)$ plane at $b_2 = 3$, $p = 2$. (b) Profile of twisted soliton in the uniform medium supported by twisted lattice soliton at $b_1 = 3.8$, $b_2 = 2$, $p = 4$, and (c) areas of stability and instability (shaded) for such solitons on the $(b_1, p)$ plane at $b_2 = 2$. (d) Stable propagation of soliton depicted in (b) in the presence of white noise. (e) Profile of three-humped soliton in the uniform medium supported by twisted lattice soliton at $b_1 = 3.57$, $b_2 = 2$, $p = 4$, and (f) its stable propagation in the presence of white noise. In (d) and (f) only second component is shown. Focusing medium $\sigma = -1$. All quantities are plotted in arbitrary dimensionless units.